\documentclass[aps,prl,floats,showpacs,twocolumn]{revtex4}
\usepackage{graphicx,soul,color,mathrsfs,wrapfig,amsmath}

\sethlcolor{yellow}

\newcommand{\ket}[1]{| #1 \rangle}
\newcommand{\mvec}[1]{\mathbf{#1}}

\newcommand{\be}{\begin{equation}}
\newcommand{\ee}{\end{equation}}

\def \ua{{\uparrow}}
\def \da{{\downarrow}}
\def \be{\begin{equation}}
\def \ee{\end{equation}}
\def \ba{\begin{array}}
\def \ea{\end{array}}
\def \bea{\begin{eqnarray}}
\def \eea{\end{eqnarray}}
\def \nn{\nonumber}

\def \half{{1\over 2}}

\def \bQ{{\bf Q}}

\def \bv{{\bf v}}

\def \bk{{\bf k}}

\def \bx{{\bf x}}

\def \a{{\alpha}}

\def \D{{\Delta}}
\def \d{{\delta }}

\def \s{{\sigma}}

\def \av#1{{\langle#1\rangle}}
\def \ket#1{{\,|\,#1\,\rangle\,}}

\begin{document}
\title{Quenching the Superconducting State of Cuprate Compounds with Electric Currents: A Variational Study}
\author{Lilach Goren and Ehud Altman}
\affiliation{Department of Condensed Matter Physics, The Weizmann Institute of Science, 76100 Rehovot (Israel)
}
\date{\today}
\begin{abstract}
We investigate the properties of cuprate superconductors subject to applied current, using modified Gutzwiller projected d-wave BCS states. The parent states include quasiparticle and quasihole pockets, of variationally determined size, generated by the current. We identify two different mechanisms for the destruction of superconductivity at the critical current: at high hole doping ($x \gtrsim 0.15$) the pockets grow and completely destroy the gap, in a BCS-like mechanism; in the underdoped regime, the superfluid stiffness vanishes at a maximal phase twist with pairing still intact. This result is indicative of a pseudogapped `normal' state which retains pairing correlations. The critical current as a function of doping displays a dome shape, similar to $T_c$. We predict unique signatures of the current induced Fermi pockets that can be seen in angle resolved photo emission spectroscopy.
\end{abstract}
\maketitle

\begin{figure}
\begin{center}
\includegraphics{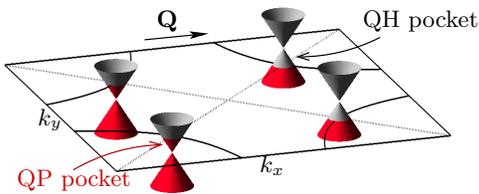}
\caption{\small  The low energy quasiparticle spectrum of a $d$-wave superconductor with a Doppler shift due to a superflow
of wave-vector $\bQ=Q\hat\bx$. The ground state consists of the marked quasiparticle and quasihole pockets.}
\label{fig:supercurrent}
\end{center}
\end{figure}
\emph{\textbf{Introduction~--}}
Shortly after the discovery of high temperature superconductivity in the cuprates, Anderson
proposed that the strong on-site repulsion and proximity to the
Mott insulator play a crucial role in determining the electronic correlations in these materials \cite{Anderson:1987}.
Such correlations are naturally included in
Gutzwiller projected variational BCS wave-functions, which indeed provide quite a good description of the superconducting state \cite{comment-Giamarchi,Paramekanti:2004}. However, variational studies have not addressed the key open questions concerning the destruction of superconductivity
with temperature, the nature of the seemingly non Fermi-liquid normal state, the pseudogap phenomenon, possible competing orders, and an associated quantum critical point.

Here we take a step toward addressing these issues using
Gutzwiller projected states. The idea is, rather than investigate the effects of temperature, which are inaccessible
in this approach, to study the consequences of an applied current at zero temperature. Both mechanisms similarly lead to the eventual demise of superconductivity and the establishment of a normal state. The reduction of the superfluid stiffness $\rho_s(T)$
is linear in temperature at low $T$ due to thermally excited quasiparticles in the $d$-wave gap nodes \cite{Lee:1997}. A similar linear reduction is expected to occur with external-current due to formation of quasiparticle (QP) pockets around the nodes, which are Doppler-shifted away from zero energy in the presence of superflow \cite{Ioffe:2002} (see illustration in Fig. \ref{fig:supercurrent}). Our approach can therefore lend insight into the
puzzle concerning the evolution of the slope $\partial\rho_s/\partial T$ with hole doping \cite{boyce:2000,comment-Lee}.
Furthermore,
the behavior of the superconductor with uniform current is intimately connected to the effects
of a magnetic field, which generates supercurrents around the vortex cores. The critical current
which we calculate
is related to the maximal circulating current found at the edge of a vortex core and the
state formed at the critical current is connected to the normal state in the core.

Whether superconductivity is suppressed by temperature, uniform current or (orbital) magnetic field, there appears
to be a qualitative change in the mechanism of its destruction and in the corresponding `normal' state,
upon doping the system from the underdoped to the overdoped regime \cite{comment-Lee}. A common view is that
in the underdoped regime superconductivity is destroyed with pairing still intact\cite{Uemura:1989,Emery:1995}, giving rise to a pseudogap above $T_c$ \cite{comment-PG}. In the overdoped regime by contrast, the transition is perceived to be conventional BCS-like. 
We shall directly test such hypotheses using the variational approach by suppressing superconductivity with a current.

To allow for the presence of a supercurrent we extend the commonly used family
of Gutzwiller projected $d$-wave BCS variational states. The current is embodied in Fermi pockets
containing the quasiparticles and quasiholes generated by the Doppler shift around the gap nodes.
A new variational parameter determines the size of the pockets for a given superflow.
We use these variational states to compute the critical current and discern the different mechanisms
that lead to destruction of superconductivity at varying hole doping, as described in the abstract.

\begin{figure}[!t]
\begin{center}
\includegraphics{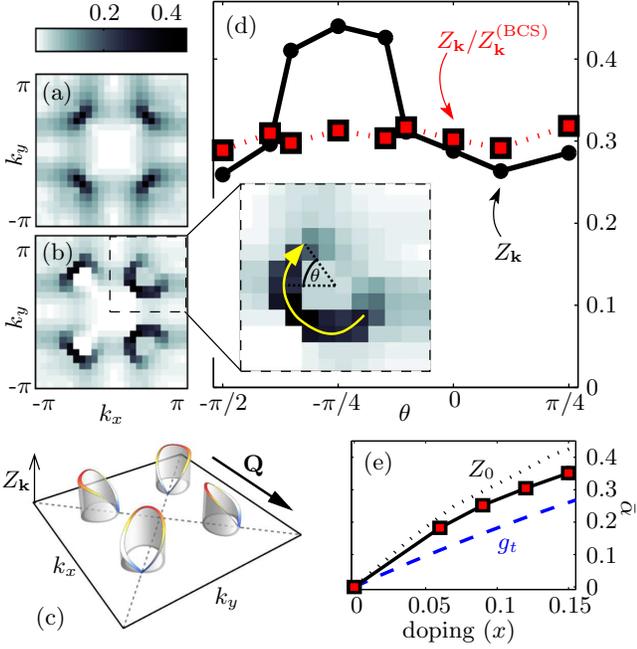}
\end{center}
\caption{{\em Discontinuities in $\langle n_{\bf k}\rangle$ and QP weight. } (a-b) Gray-scale plots of $ Z_{\bf k}\equiv|\nabla_{\bf k} n_{\bf k}|$ on a $24\times 24$ lattice, doping $x=0.12$ and $Q=0$ (a) and $Q=\pi/2$ (b). (c) The QP weight on the current-induced Fermi surfaces in a $d$-wave BCS state. (d) QP weight $ Z_{\bf k}(Q=\pi/2)$ (black circles) and the QP weight renormalization $\alpha_{\bf k}\equiv Z_{\bf k}/ Z_{\bf k}^{({\rm BCS})}$ (squares)
along the quasihole pocket Fermi surface (see inset). (e) Fermi surface average of ${\alpha_{\bf k}}$ as a function of doping (squares) compared with the zero current result of Ref. \cite{Paramekanti:2004} (dotted line) and the renormalized mean field result $g_t=2x/(1+x)$ (dashed line).
}
\label{fig:gradnk}
\end{figure}

  \emph{\textbf{The variational wave-function~--} }
We set out to investigate the ground state of the $t$-$t'$-$J$ Hamiltonian on a square lattice subject to superflow.
In order to later fix the superflow wave-vector $\bQ=Q\hat{x}$ it is convenient to impose a constant
vector potential along the $\hat{\bx}$ direction via a Peierls substitution,
\[
H(\mvec{Q})=P_G\sum_{ i,j,\sigma}t_{ij}e^{\frac{i}{2}Q(x_j-x_i)} c_{i\sigma}^\dag c_{j\sigma}P_G+h.c +J \sum_{\langle ij \rangle}\mvec{s}_i\mvec{s}_j.
\]
Here $P_G=\Pi_i(1-n_{i\ua}n_{i\da})$ is the Gutzwiller projection operator.
The variational wave-functions we use are Gutzwiller projected $d$-wave BCS states containing
filled QP and quasihole (QH) pockets around the nodes. Besides  the usual
parameters $\Delta$ and $\mu$ we include a new variational parameter $Q_{var}$
which determines the size of the current-induced pockets. The superflow is imposed
by constraining $\D$ to be spatially uniform (and real) regardless of $\bQ$.

To motivate the construction of the wave-function, it is useful to recall the mean field BCS
description of a
$d$-wave superconductor subject to current of wave-vector $\bQ$. The Bogoliubov De Gennes
spectrum is given by
 \bea
 \label{SpectrumwCurrent}
 E_{\bf k}^{\pm}({\bf Q})&=&\xi_{-,{\bf k}}(\mvec{Q}) \pm\sqrt{{\xi}^2_{+,{\bf k}}(\mvec{Q})+\Delta_{\bf k}^2}\nn\\
 &\approx&\frac{1}{2}\mvec{v}_{\text{F}}(\bk)\cdot \mvec{Q}\pm E_\bk(0)
  \eea
where $\xi_\bk$ is the single particle dispersion, $\D_\bk=\D(\bQ)(\cos k_x \!-\!\cos k_y)$ the $d$-wave gap function, and
$\xi_{\pm,\bf k}\equiv  (\xi_{{\bf k}+\mvec{Q}/2}\pm\xi_{{\bf k}-\mvec{Q}/2})/2$. The second line above gives the low energy spectrum
for small $\bQ$.
The BCS ground state is defined by occupation of all the negative energy QP states.
Because of the Doppler shift $\bv_{\text{F}}\cdot\bQ/2$,
this implies filling an elliptical pocket of QPs around the two downstream nodes
and QHs around the upstream nodes, cf. Fig.~\ref{fig:supercurrent}.
More generally, due to interactions, the Doppler shift may be renormalized to
$\a \bv_{\text{F}}\cdot\bQ/2\!\equiv\! \bv_{\text{F}}\cdot\bQ_{var}/2$. The parameter $\a$ is commonly referred to as
the effective QP current renormalization \cite{Lee:1997,Nave:2006}.
In the parent BCS state we shall allow the freedom
to determine the size of the pocket via the additional variational parameter $Q_{var}$ independent of the externally
applied twist $\bQ$.

The variational state we construct is a Gutzwiller projection of the mean field ground state with $N$ electrons,
  \[
\label{ }
\ket{\Psi^{(N)}[\Delta,\mu,Q_{\rm var}]}=P_G P_N  \prod_{\sigma,\mvec{k}\in \mathcal{P}(Q_{\rm var})}    \tilde{\gamma}^\dagger_{\mvec{k} \sigma}\ket{\Psi_{\rm{BCS}}({\bf Q})}.
\]
$P_N$ projects the state onto the N-particle subspace, $\mathcal{P}(Q_{\rm var})$ denotes the wave-vectors inside the QP pockets, and $\ket{\Psi_{\rm{BCS}}({\bf Q})}$ is the vacuum of the QP operators,
 \begin{equation}
\label{ }
\tilde{\gamma}^\dagger_{k\sigma}=\tilde{u}_k({\bf Q})c_{k\sigma}^\dagger -\sigma \tilde{v}_k({\bf Q})c_{-k\bar{\sigma}}
\end{equation}
with
$
\tilde{u}_{\bf k}^2({\bf Q}) = 1-\tilde{v}_{\bf k}^2({\bf Q}) =\half \left[1+{\xi}_{+,\bf k}/({{\xi}_{+,\bf k}^2+\D_\bk^2})^{1/2}\right] $.

We minimize the energy as a function of the variational parameters $\Delta,\mu$ and $Q_{var}$ using
variational Monte Carlo (VMC) \cite{comment-Giamarchi,Paramekanti:2004}, with $J=t/3$ and $t'=-t/4$.
  The optimized energy and wave-function are then used to study the effects of strong correlations on a current-carrying $d$-wave superconductor.

 \emph{\textbf{Quasiparticle properties of the projected state~--} }
 We turn to investigate basic properties of the projected current-carrying state as would be observed in angle resolved photoemission spectroscopy (ARPES). 
 Specifically we consider $\av{n_\bk}$, related to the spectral function via the sum rule
 $\langle n_{{\bf k}}\rangle=\sum_{\s}\langle c_{{\bf k}\s}^{\dag}c_{{\bf k}\s}\rangle=\int_{-\infty}^0\!\! d\omega A_{\bf Q}({\bf k},\omega)$. From the variational state we can extract discontinuities in $\langle n_{{\bf k}}\rangle$ which correspond to the QP weights  at $\omega\!=\!0$.

To study the effects of strong correlations we compare the structure of $\langle n_{{\bf k}}\rangle$ in the projected state with that of the underlying BCS wave-function. Within BCS, the $T=0$ spectral function has the form
 \be
\label{AwithTwist:BCS}\nn
A_{\bf Q}^{\rm BCS}({\bf k},\omega)=|\tilde{u}_{\bf k}|^2\delta[\omega-E_{\bf k}^{+}({\bf Q})]+ |\tilde{v}_{\bf k}|^2\delta[\omega-E_{\bf k}^{-}({\bf Q})]
\ee
 Consequently, $\langle n_{{\bf k}}\rangle$ has discontinuities of magnitude $\tilde{u}_{\bf k}^2$ ($\tilde{v}_{\bf k}^2$)  along the QP (QH) pocket Fermi surfaces, as depicted in Fig.~\ref{fig:gradnk}(c). At $Q=0$ the pockets shrink to 4 nodal points with a QP weight of 1 along the nodal direction at each point\cite{comment-berg}.

 For the comparison with BCS, we plot  $|\nabla_{\bf k}n_{\bf k}|\equiv  Z_{\bf k}$ in the projected state with ${\bf Q}=0$ and ${\bf Q}=\pi/2\  \hat{x}$ [Fig.~\ref{fig:gradnk}(a,b)].
 The pocket Fermi surfaces in Fig.~\ref{fig:gradnk}(b), marked by the peaks of $  Z_{\bf k}$, behave similarly to the BCS case shown in Fig.~\ref{fig:gradnk}(c). To quantify the relation we extract the ratio
 between the QP weight after projection and the bare BCS value along the hole-pocket boundary
 $\alpha_{\bf k}\equiv Z_{\bf k}/ Z_{\bf k}^{\rm BCS}$ [filled squares in Fig.~\ref{fig:gradnk}(d)]. Interestingly,
 $\alpha_{\bf k}$ is almost independent of ${\bf k}$. This is consistent with the central assumption of
  renormalized mean field theories (RMFT) \cite{Zhang:1988,Kotliar:1988}, that the
  Gutzwiller projection can be encapsulated in a global renormalization factor.
   We also find that the renormalization factor is independent of current, and its doping dependence (Fig.~\ref{fig:gradnk}(e)) matches with the zero current result obtained in Ref. \cite{Paramekanti:2004}.

The above predictions for ARPES do not depend on the pocket shape. For clarity, we chose $\D$, $\mu$ and $Q_{\rm var}$ to obtain circular pockets rather than the variationally determined elliptical shape.

\begin{figure}[!t]
\begin{center}
\includegraphics{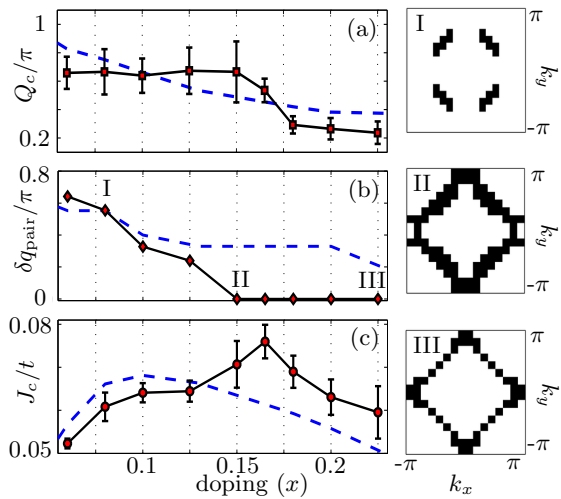}
\end{center}
\caption{ {\em Critical flow vs doping} calculated from VMC (solid lines) vs RMFT (dashed lines) results. (a) Critical flow wave-vector $Q_c$. (b) Length of residual paired region in the underlying Fermi surface. (c) Critical current $J_c$ vs doping. Right panels: pockets at the critical flow for three doping levels marked in panel (b). The pockets grow with doping until they touch the zone boundary at $x\ge0.15$.}
\label{fig:QcJcAk}
\end{figure}

 \emph{\textbf{Critical current in the variational state~--}}
 The critical current is the maximal current that can be imposed on a superconductor
  \be\label{Jc}
 J_c=\max_{Q}\frac{dE}{dQ}
 \ee
 and it occurs at a critical flow wave-vector $Q=Q_c$. The maximum is taken under the constraint that at the flow $Q$ the variational Fermi pockets do not yet cover the underlying large Fermi surface, at which point superconductivity would be lost due to complete depairing.
  Technically, we obtain $Q_c$ and $J_c$ by fitting the variational energy, calculated on a $18\times 18$ lattice, to a polynomial \cite{comment-fit}
  \be\label{EQvar}
 E(Q)= E_0 + \frac{1}{2}\rho_{s0} {Q}^2 -\frac{1}{6}\eta {|Q|}^3 +O(Q^4).
 \ee

The results of this analysis are shown in Fig. \ref{fig:QcJcAk}.
Figure \ref{fig:QcJcAk}(a) displays the variationally determined $Q_c$ as a function of the doping $x$. Notably, $Q_c$ is essentially independent of $x$ for  $x\le x_0=0.15$, and it drops sharply for higher doping.
A complementary viewpoint is given in Fig.~\ref{fig:QcJcAk}(b) showing $\d q_{\rm pair}$, the length of the residual gapped region of the Fermi surface, at the critical current. As doping is increased, the maximal pocket size grows and $\d q_{\rm pair}$ drops, until it reaches zero at $x=x_0$. At $x\ge x_0$ the critical current is obtained when the wave-function is fully depaired. The Fermi pockets at the critical current are shown in panels I-III in Fig.~\ref{fig:QcJcAk} for three doping levels.

We conclude that optimal doping marks a transition between two distinct mechanisms for
the destruction of superconductivity with current: A BCS-like mechanism for $x\!>\!x_0$, and a "bosonic" mechanism, whereby superfluidity is destroyed with pairing still present, for $x\!<\!x_0$. The doping independent value of $Q_c$ suggests an analogy with lattice bosons for which
superfluidity is destroyed at $Q_c\!=\!\pi/2$ \cite{Polkovnikov:2005}.
The residual pairing at the critical current suggests that the normal state formed upon destruction of superconductivity in the underdoped regime retains pairing correlations.

The transition between the two mechanisms gives rise to a maximum of $J_c$ at $x\simeq x_0$\cite{comment-peak}. The dome shape of the critical current as a function of doping, shown in Fig.~\ref{fig:QcJcAk}(c), is reminiscent of the observed doping dependence of the critical temperature, with a similar optimal doping level. We can estimate the expected critical current in the underdoped regime as $J_c\sim \rho_{s0}Q_c$. Taking $\rho_{s0}=\rho_{ab}$ from experiments, such as \cite{boyce:2000} and $Q_c\sim\pi/2$ gives a (3D) current density of $J_c/d \sim 10^{7} {\rm A/cm^2}$, where $d$ is the interlayer spacing. This is an estimate for the microscopic critical current, or the maximal current at the edge of a vortex core.
The macroscopic critical current may be significantly suppressed due to inhomogeneity. 

Before proceeding we note, that similar behavior might have been expected within RMFT \cite{Zhang:1988}, due to the competition between the rising stiffness and falling gap which was noted early on \cite{Kotliar:1988}. Nevertheless, a calculation of the critical flow within RMFT (dashed lines in  Fig.~\ref{fig:QcJcAk}) draws a markedly different picture than does the variational calculation.
 In particular the critical flow wave-vector drops continuously as $1/x^2$ and the pocket size is always smaller than the Brillouin zone, so that there is no complete depairing at any doping level within RMFT.

 \emph{\textbf{The current dependent superfluid stiffness~-- }}The current dependent superfluid stiffness is the nonlinear response $\rho_s(Q_0)=\partial^2_Q E|_{Q=Q_0}$
 and is calculated within the variational scheme by fitting the optimized energy to Eq. (\ref{EQvar}) in the range $0\!<\!Q\!<\!Q_c$ \cite{comment-fit}.

\begin{figure}[!t]
\begin{center}
\includegraphics{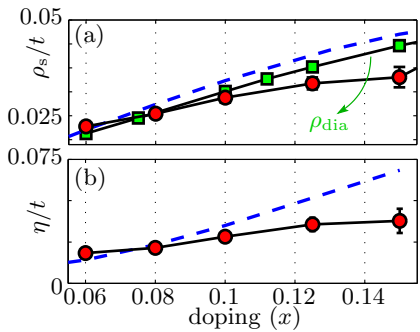}
\end{center}
\caption{ (a) Zero current superfluid stiffness $\rho_{s0}$ vs hole doping in the underdoped regime (circles), compared with the diamagnetic response (squares), computed without allowing for pockets, and with the RMFT result for $\rho_{s0}$ (dashed). (b) The slope $\eta=\partial\rho_s/\partial Q$ vs doping. Compare with the much weaker doping dependence in
the RMFT result (dashed).}
\label{fig:response}
\end{figure}

The result is displayed in Fig~\ref{fig:response}(a), showing the doping dependence of the variational stiffness $\rho_{s0}$ (solid circles) compared with the mean field calculation (dashed). Note that we obtain an improved estimate of the superfluid stiffness, including paramagnetic contributions, compared with the diamagnetic upper bound (squares) that was calculated previously \cite{Paramekanti:1998}.

More importantly, we can now study the suppression of superfluidity with the flow, which
at small $\bQ$ has the form
 $\rho_s(Q)\simeq \rho_{s0}-\eta|Q|$. Figure ~\ref{fig:response}(b) displays the coefficient $\eta$ as a function of doping extracted from the variational calculation (solid circles).
 This doping dependence is clearly weaker than the RMFT result (dashed)
 in which $\eta \sim x^3$ \cite{Ioffe:2002}.
 We note that in the underdoped regime $\eta$ can be related to $Q_c$ via (\ref{EQvar})
 as $\eta\simeq\rho_{s0}/Q_c$. Since $\rho_{s0}\sim x$ , the weak doping dependence
  of $\eta$ can be traced back to the doping independent
 critical flow $Q_c$. Hence it also stems from the bosonic mechanism of destruction of superconductivity in the underdoped regime.

It is plausible that the weak doping dependence of
$\partial\rho_s/\partial Q$ shares the same origin as the unexplained weak doping
dependence of $\partial\rho_s/\partial T$ \cite{boyce:2000}. The latter is not understood within
mean field approaches\cite{Lee:1997,comment-Lee} nor from calculations of the current
carried by a Gutzwiller projected single QP state \cite{Nave:2006}.
Both slopes stem from paramagnetic current of QPs near the nodes, induced either by temperature or current. Measurements of the stiffness reduction in the presence of supercurrent may clarify the relation between $\partial \rho_{s}/dT$ and $\partial \rho_{s}/dQ$ in cuprates.

 \emph{\textbf{Summary~--}}
Using a new class of variational states we identified a transition between two mechanisms for the destruction of superconductivity with current. In the overdoped regime ($x>0.15$) we found a BCS-like mechanism, where the critical current is reached due to complete depairing, while in the underdoped regime we observed a bosonic mechanism where superconductivity is destabilized while pairing survives.
The critical current has a dome shape with an optimal doping set by this transition \cite{comment-Ganesh}. 
We computed the linear reduction of the superfluid stiffness
at small current. The slope $d\rho_s/dQ$ depends only weakly on doping, 
reflecting the puzzlingly weak doping dependence of $d\rho_s/dT$ in
experiments. 

Though our results pertain to the destruction of superconductivity with current at zero temperature, they are strongly
suggestive of a pairing mechanism underlying the pseudogap phenomenology in the underdoped cuprates at $T\!>\!T_c$.
Further insight into the nature of the normal state can be gained by extending our work to allow
competing orders, such as spin or charge-density wave, possibly reducing the variational energy
as superconductivity is suppressed \cite{Zhang:2002,Shay:2009}. This may also elucidate the  role played by a quantum critical point, where the competing order in the normal state vanishes \cite{Panagopoulos:2002}.

We thank A. Paramekanti, A. Kanigel, A. Auerbach, and S.D. Huber for stimulating discussions.
This work was supported in part by grants from the Israeli Science Foundation and the Minerva foundation.


\begin{thebibliography}{10}

\bibitem{Anderson:1987}
P.W.\ Anderson,
 Science {\bf 235}, 1196 (1987).

\bibitem{comment-Giamarchi}
T.\ Giamarchi and C.\ Lhuillier, Phys.\ Rev.\ B {\bf 43}, 12943 (1991); B.\
  Edegger, V.\ Muthukumar, and C.\ Gros, Adv.\ in Phys.\ {\bf 56}, 927 (2007);
  S.\ Yunoki, E.\ Dagotto, and S.\ Sorella, Phys.\ Rev.\ Lett.\ {\bf 94},
  037001 (2005).

\bibitem{Paramekanti:2004}
A.\ Paramekanti, M.\ Randeria, and N.\ Trivedi,
 Phys.\ Rev.\ B {\bf 70}, 054504 (2004).

\bibitem{Lee:1997}
P.A.\ Lee and X.G.\ Wen,
 Phys.\ Rev.\ Lett.\ {\bf 78}, 4111 (1997).

\bibitem{Ioffe:2002}
L.B.\ Ioffe and A.J.\ Millis,
 J.\ Phys.\ Chem.\ Solids {\bf 63}, 2259 (2002).

\bibitem{boyce:2000}
B.R.\ Boyce, J.\ Skinta, and T.R.\ Lemberger,
 Physica C {\bf 341-348}, 561 (2000).

\bibitem{comment-Lee}
P.A.\ Lee, N.\ Nagaosa, and X.G.\ Wen, Rev.\ Mod.\ Phys.\ {\bf 78}, 17 (2006)
  and references therein.

\bibitem{Uemura:1989}
{Y.J.Uemura et al.},
 Phys.\ Rev.\ Lett.\ {\bf 62}, 2317 (1989).

\bibitem{Emery:1995}
V.J.\ Emery and S.A.\ Kivelson,
 Nature {\bf 374}, 434 (1995).

\bibitem{comment-PG}
See e.g H.\ Ding et al., Nature {\bf 382}, 51 (1996).

\bibitem{Nave:2006}
C.P.\ Nave, D.A.\ Ivanov, and P.A.\ Lee,
 Phys.\ Rev.\ B {\bf 73}, 104502 (2006).

\bibitem{comment-berg}
To compare with ARPES we should plot the gauge invariant quantity $\langle
  n_{{\bf k-Q}/2}\rangle$.

\bibitem{Zhang:1988}
F.\ Zhang, C.\ Gros, T.M.\ Rice, and H.\ Shiba,
 Supercond.\ Sci.\ Technol.\ {\bf 1}, 36 (1988).

\bibitem{Kotliar:1988}
G.\ Kotliar and J.\ Liu,
 Phys.\ Rev.\ B {\bf 38}, 5142 (1988).

\bibitem{comment-fit}
When fitting we keep terms up to $O(Q^5)$.

\bibitem{Polkovnikov:2005}
A.\ Polkovnikov {\em et~al.},
 Phys.\ Rev.\ A {\bf 71}, 063613 (2005).

\bibitem{comment-peak}
In the BCS-like regime $J_c(x)\sim \rho_s(x)\Delta(x)/v_f $ is maximized at $x$
  slightly larger than $x_0$.

\bibitem{Paramekanti:1998}
A.\ Paramekanti, N.\ Trivedi, and M.\ Randeria,
 Phys.\ Rev.\ B {\bf 57}, 11639 (1998).

\bibitem{comment-Ganesh}
A similar transition, with a more obvious origin is found in the negative $U$
  Hubbard model upon going from weak to strong coupling in R.\ Ganesh, A.\
  Paramekanti, and A.A.\ Burkov, Phys.\ Rev.\ A {\bf 80}, 043612 (2009).

\bibitem{Zhang:2002}
Y.\ Zhang, E.\ Demler, and S.\ Sachdev,
 Phys.\ Rev.\ B {\bf 66}, 094501 (2002).

\bibitem{Shay:2009}
M.\ Shay {\em et~al.},
 Phys.\ Rev.\ B {\bf 80}, 144511 (2009).

\bibitem{Panagopoulos:2002}
C.\ Panagopoulos {\em et~al.},
 Phys.\ Rev.\ B {\bf 66}, 064501 (2002).

\end{thebibliography}
\end{document}